\documentclass[pra,twocolumn,amsmath,amssymb]{revtex4-1} %

\usepackage{epsfig,amsmath}
\usepackage{subfigure}
\usepackage{graphicx}
\usepackage{dcolumn}
\usepackage{stmaryrd}
\usepackage{mathrsfs}
\usepackage{pifont}
\usepackage{amsthm}
\usepackage{amssymb}
\usepackage{bm}
\usepackage{latexsym}
\usepackage[colorlinks=true,linkcolor=blue,citecolor=blue]{hyperref}
\usepackage{color}

\theoremstyle{plain}

\begin{document}

\title{Exactly solvable pairing models in two dimensions}

\author{Lian-Ao Wu}

\affiliation{Department of Theoretical Physics and History of Science, The Basque Country
University (EHU/UPV), P. O. Box 644, E-48080 Bilbao, Spain and IKERBASQUE,
Basque Foundation for Science, E-48011 Bilbao, Spain}

\date{\today}

\begin{abstract}
The BCS theory models electron correlations with pure zero-momentum pairs. Here we
consider a family of pairing Hamiltonians, where the electron correlations
are modelled with pure arbitrary-momentum pairs. We find
all models in the family are exactly solvable, and present these solutions. It is interesting to note that the $\eta$ pair or the $d$ -wave pair condensate in 
$T_{c}$ superconductivity can be the ground state of a Hamiltonian in the family. These models are two-dimensional  because only the z-component of the total electron
spin $S_z$ is conserved. Significantly, for the $\eta$ pair or $d$ -wave pairing model in the family we find an analytical expression of energy and an abrupt ground state change from independent particle state to the $d$ -wave pair condensate, suggesting a quantum phase transition. 
\end{abstract}

\maketitle

\section{Introduction}
In the conventional BCS theory,  the zero-momentum pairing is assumed to be dominant in the electron-electron interaction~\cite{BCS,Anderson93}. 
For degenerate single-particle energies, the model becomes analytically solvable and leads to a BCS pairing condensate as the ground state, which provides straightforward explanations for many physical phenomena. Over the past decades,  pure nonzero-momentum pairing has attracted much attentions from physicists, for instance, the $\eta$ pairing  \cite{Yang} as metastable eigenstates of a simple Hubbard model and the $d$-wave pairing related to antiferromagnetic spin and unconventional high $T_c$ superconductivity~\cite{Monthoux92}. Therefore, motivated by the solvable BCS model, it is interesting to ask: what are the Hamiltonian models that lead to these nonzero-momentum condensates as the ground states?  This is, we wish to reveal the dynamics of creation of the condensates and find applications, for instance by inversely knowing these Hamiltonians it might be possible to inversely implement the $d$-wave pairing condensates to engineer unconventional high $T_c$ superconductivity with state-of-the-art technologies, such as trapped ions.  

In this paper, we start with a general electron Hamiltonian and consider a
family of pairing models, including those with the BCS pairs \cite{BCS}, the 
$\eta $ pairs \cite{Yang} and the $d$-wave pairs \cite{SO5,SO5Wu} in $T_{c}$
superconductivity. We find exact solutions to this family, with the
assistance of the dressing transformations introduced for qubits \cite{Wu03}%
. The total electron spin $\overrightarrow{S}$ of these models are not
conserved except for the BCS model, although the conservation of the $z$%
-component $S_{z}$ of the total spin remains. These models are legitimate as
physical Hamiltonians only on an $xy$ - plane defined by $S_{z}$. 
Different pairings in these models are on equal footing in the sense that
they can be unitarily transformed into each other. In particular, these
models can be converted to the conventional BCS model, with an interesting
extra single-particle energy. We study this extra contribution of
single-particle energy and observe that it plays a decisive role in a
quantum phase transition between independent-particle ground states and collective $\eta $
-pairing or $d$-wave pairing ground states. We find conditions where non-BCS
pairs, including the $\eta $-pairs and the $d$-wave pairs, play the same
role as the BCS pairs. In this case, the ansatz in constructing the BCS
model implies that these non-BCS pairs could become dominant.

\section{General electron Hamiltonian, pairs with given momentum and SU(2) symmetry}

Consider a general Hamiltonian of electrons on a
periodic lattice. 
\begin{equation}
H=\sum_{k,\sigma =\downarrow ,\uparrow }\epsilon _{k}n_{k\sigma }+V,
\label{general}
\end{equation}%
where $n_{k\sigma }=c_{k\sigma }^{\dagger }c_{k\sigma }$ is the number\
operator of an electron with momentum vector $k$ and spin $\sigma $. The
electron-electron interaction is 
\begin{equation}
V=\sum_{q,k,k^{\prime }}V_{k,k^{\prime }}c_{q-k^{\prime }\downarrow
}^{\dagger }c_{k^{\prime }\uparrow }^{\dagger }c_{k\uparrow
}c_{q-k\downarrow }.  \label{interaction}
\end{equation}%
Here $c_{k\uparrow }(c_{k\downarrow })$ is a momentum-space annihilation
operator of a spin-up (spin-down) electron. We consider one-, two- or
three- dimensional cubic lattices, with the total number $L$ of lattice sites in each
dimension. The vectors or modes $k=$ $(k_{x},k_{y},k_{z})$ with $%
k_{x,y,z}=2\pi l/L$, where $l=0,1,...,L-1$. Vectors $q$ have the same modes
as $k$. The interaction can be rewritten as 
\begin{equation}
V=\sum_{q,k,k^{\prime }}V_{k,k^{\prime }}\eta _{q}^{\dagger }(k^{\prime
})\eta _{q}(k),  \label{pairforce}
\end{equation}%
which is expressed in terms of pair operators 
\begin{equation}
\eta _{q}(k)=c_{k\uparrow }c_{q-k\downarrow }\text{ and }\eta _{q}^{\dagger
}(k)=c_{q-k\downarrow }^{\dagger }c_{k\uparrow }^{\dagger }.  \label{qpair}
\end{equation}%
It is easy to check that, for a given $q$, these operators satisfy
commutation relations, 
\begin{equation}
\lbrack \eta _{q}(k),\eta _{q}^{\dagger }(k^{\prime })]=\delta _{kk^{\prime
}}(1-c_{k\uparrow }^{\dagger }c_{k\uparrow }-c_{q-k\downarrow }^{\dagger
}c_{q-k\downarrow }),  \label{qpairC}
\end{equation}%
such that $\eta _{q}(k),\eta _{q}^{\dagger }(k)$ and $[\eta _{q}(k),\eta
_{q}^{\dagger }(k)]$ close an $su(2)$ algebra. In other words, they play the
same roles as Pauli matrices\cite{Wu02} , 
\begin{eqnarray}
\sigma _{k}^{-} &\Longleftrightarrow &\eta _{q}(k),  \label{Pauli} \\
\sigma _{k}^{+} &\Longleftrightarrow &\eta _{q}^{\dagger }(k),  \nonumber \\
\sigma _{k}^{z} &\Longleftrightarrow &(c_{k\uparrow }^{\dagger }c_{k\uparrow
}+c_{q-k\downarrow }^{\dagger }c_{q-k\downarrow }-1)/2,  \nonumber
\end{eqnarray}%
acting on the two bases $\left| 0\right\rangle _{k}=I_{k}\left|
0\right\rangle $ and $\left| 1\right\rangle _{k}=\eta _{q}^{\dagger
}(k)\left| 0\right\rangle $, where $\left| 0\right\rangle $ is the vacuum
state and $I_{k}$ is a unit operator on the $k$-th mode. These define a
qubit, or precisely the $k$-th qubit in a \emph{collective} subspace of the
entire electron Hilbert space.

In case that $V_{k-k^{\prime }}=G$ is constant, the interaction $
V=G\sum_{q,k,k^{\prime }}\eta _{q}^{\dagger }(k^{\prime })\eta _{q}(k)$
corresponds to a one-band Hubbard-like model \cite{Freericks}, which is a
sum of different \emph{q-components} $\sum_{k^{\prime }}\eta _{q}^{\dagger
}(k^{\prime })\sum_{k}\eta _{q}(k)$ (Note that here $q$ is not summed over).
In contrast, the BCS ansatz considers particularly the $q=0$ component and
also sets $V_{k,k^{\prime }}=G$, such that $V=G\sum \eta _{0}^{\dagger
}(k^{\prime })\eta _{0}(k)$. Here $G$ is negative for the BCS model.

\section{Simplified interactions: pure pairing models}

Generally, pairs $%
\eta _{q}(k)$ and $\eta _{q^{\prime }}^{\dagger }(k)$ for different values
of $q$ do not commute but are related to each other. As an ansatz, the BCS theory chooses
the simplest $q=0$ pair component and neglects the others, which has been
verified by numerous experiments. Motivated by this BCS ansatz,
we consider the family of all pure \emph{q-} components in the general
Hamiltonian (\ref{general}). We also employ a sightly general separable
couplings $V_{k,k^{\prime }}=Gg(k)g(k^{\prime }),$ where the BCS assumption $%
V_{k,k^{\prime }}=G$ \ is a special case when $g(k)=g(k^{\prime })=1$. The
family of pure $q$ -component models can therefore be written as 
\begin{equation}
H_{q}=\sum_{k,\sigma =\downarrow ,\uparrow }\epsilon _{k}n_{k\sigma }+V_{q},
\label{qHamiltonian}
\end{equation}%
where $V_{q}=G\eta _{q}^{\dagger }\eta _{q}$ and $q$ runs over the total
momentum space. This paper will mostly concentrate on the cases with $G<0$
as in the BCS theory. We term operators $\eta _{q}$ as $q$-pairs, which are
defined as 
\begin{eqnarray}
\eta _{q} &=&\sum_{k}g(k)c_{k\uparrow }c_{q-k\downarrow },
\label{collective} \\
\eta _{q}^{\dagger } &=&\sum_{k}g(k)c_{q-k\downarrow }^{\dagger
}c_{k\uparrow }^{\dagger },
\end{eqnarray}%
behaving as \emph{collective} pairs. We also call the pure $q$-component
models $H_{q}$ as pure $q$-pairing models. The pair with $q=0$ and $g(k)=1$
corresponds to the BCS collective pair.

It is interesting to note that the $\eta _{\pi }$ pair with $g(k)=1$ is the $%
\eta $ pair introduced in \cite{Yang}, but it corresponds to the $d$-wave
pair \cite{SO5,SO5Wu} when 
\begin{equation}
g(k)=\text{sign}(\cos k_{x}-\cos k_{y})=\pm 1,  \label{gk}
\end{equation}
recommended in ref. \cite{Henley}. In order for $\pi -k$ and $k$ to be
simultaneously possible $k$ values, the number of sites, $L,$ must be 
\textit{even} for the $\eta $ pair \cite{Yang}.  We apply this constraint to
arbitrary value of $q=2\pi /L$ (three-dimensional integer) (mod$2\pi )$,
requiring that the even-odd parity of $L$ is the same as that of $2\pi /q$,
otherwise $q-k$ and $k$ would not be simultaneously possible $k$ values.

\section{Symmetry constrains}
The total spin operator of electrons is 
\[
\vec{S}=\frac{1}{2}\sum_{k\alpha \beta }c_{k\alpha }^{\dagger }\vec{\sigma}%
_{\alpha \beta }c_{k\beta }, 
\]%
where $\vec{\sigma}=(\sigma _{x},\sigma _{y},\sigma _{z})$ are Pauli's
matrices. The BCS model $H_{0}$ conserves the total spin $\vec{S}$, i.e., $%
[H_{0},\vec{S}]=0$. However, while $S_{z}$ remains conserved, the total spin
is no longer conserved for the rest of $q\neq 0$ pairing models (\ref%
{qHamiltonian}). It indicates that $H_{q\neq 0}$ works in an anisotropic
spin space, or a two-dimensional $xy$-plane perpendicular to the $z$-axis in
spin space. Quantum states are common eigenvectors of $S_{z}$ and $H_{q\neq
0}$.  We have not found further coordinate space symmetry for $H_{q\neq 0}$ ,
however an assumption of electrons being on the $xy$ plane of the coordinate
space is clearly consistent with the conservation of $S_{z}$.

\section{Dressing transformation and the BCS model.}
 We now come to
the main results of this paper. Specifically,  we first introduce the
following dressing transformations \cite{Wu03} 
\begin{equation}
\mathcal{W}_{qq^{\prime }}=\exp (-\frac{\pi }{2}[\sum_{k}g(k)(c_{q-k%
\downarrow }^{\dagger }c_{q^{\prime }-k\downarrow }-c_{q^{\prime
}-k\downarrow }^{\dagger }c_{q-k\downarrow })]),  \label{dress1}
\end{equation}%
where  we set $g^{2}(k)=1$ as done in \cite{Henley} and will allow it to be an
arbitrary real function later. One can check that any two $q$-pairs, $\eta
_{q}$ pairs and $\eta _{q{\prime }}$ pairs, can be transformed or rotated
unitarily into each other, i.e., $\eta _{q{\prime }}=\mathcal{W}_{qq^{\prime
}}^{\dagger }\eta _{q}\mathcal{W}_{qq^{\prime }}$. These unitary dressing
transformations do not change the $su(2)$ commutation relation (\ref{qpairC}%
) but rotate the bases, $\eta _{q}^{\dagger }(k)\left| 0\right\rangle
\rightarrow \eta _{q^{\prime }}^{\dagger }(k)\left| 0\right\rangle $. The
forms of electron-electron interactions in $H_{q}$ are invariant under these
transformations,%
\[
V_{q^{\prime }}=\mathcal{W}_{qq^{\prime }}^{\dagger }V_{q}\mathcal{W}%
_{qq^{\prime }}. \label{main} 
\]%
It also shows that pair correlations with different $q$ are not independent.
On the contrary, these correlations are equivalent, or \emph{similar} in
mathematical term, to the BCS pair correlation $V_{0}$, subject to the
unitary transformations $\mathcal{W}_{q0}$ (notation as $\mathcal{W}_{q}$
for simplicity). Both pairs must live on the two-dimensional spin space due
to the conservation of $S_{z}$.

It is notable that under the dressing transformations (\ref{dress1}), the
forms of single-particle energies are likewise invariant,%
\[
\mathcal{W}_{q}^{\dagger }\sum_{k,\sigma =\downarrow ,\uparrow }\epsilon
_{k}c_{k\sigma }^{\dagger }c_{k\sigma }\mathcal{W}_{q}=\sum_{k,\sigma
=\downarrow ,\uparrow }\epsilon _{k\sigma }(q)c_{k\sigma }^{\dagger
}c_{k\sigma } , \label{main1} 
\]%
where $\epsilon _{k\sigma }(q)=\epsilon _{k}\delta _{\sigma \uparrow
}+\epsilon _{k-q}\delta _{\sigma \downarrow } $. It remains diagonal but
becomes spin-dependent. This concludes that the $q$-pair Hamiltonian $H_{q}$
is equivalent to the BCS Hamiltonian with a spin-dependent single-particle
energy. The spin dependence can be separated from the total single-particle
energy. Consequently, the $q$-pairing models are converted to an exactly
same form as the BCS model with spin-\emph{independent} levels $\epsilon
_{k}(q)$ plus a single-particle Hamiltonian commutable with the
electron-electron interaction. The effective levels now are functions of $q$%
. \emph{A }$q$\emph{\ -pair model can therefore be treated as a BCS model }$%
\tilde{H}_{0}$\emph{\ with single-particle levels }$\epsilon _{k}(q)$\emph{\
plus an extra }$\bar{h}$\emph{\ that commutes with }$\tilde{H}_{0}$.

\section{Exact solutions for the family of q-pair Hamiltonians.}

The equivalence or similarity leads to exact solutions of all models in this
family. The first step is to solve the eigenproblem of the BCS Hamiltonian $%
\tilde{H}_{0}=\sum_{k}\epsilon _{k}(q)(n_{k\uparrow }+n_{k\downarrow
})+G\eta _{0}^{\dagger }\eta _{0}$, 
\[
\tilde{H}_{0}\left| \Psi \right\rangle =E\left| \Psi \right\rangle, 
\]
where $\tilde{H}_{0}+\bar{h}=\mathcal{W}_{q}^{\dagger }H_{q}\mathcal{W}_{q}$ 
$\ $and $\epsilon _{k}(q)=(\epsilon _{k}+\epsilon _{k-q})/2$ are effective
single-particle levels. The extra single-particle energy $\bar{h}
=\sum_{k}(\epsilon _{k}-\epsilon _{k-q})(n_{k\uparrow }-n_{k\downarrow })$
commutes with $\tilde{H}_{0}$, such that common eigenfunctions of $\tilde{H}%
_{0}$ and $\bar{h}$ are allowed. Note that we have rewritten the
single-particle energy of $H_{q}$ as 
\begin{eqnarray*}
&&\sum_{k,\sigma =\downarrow ,\uparrow }\epsilon _{k\sigma }(q)c_{k\sigma
}^{\dagger }c_{k\sigma } \\
&=&\bar{h}+\sum_{k}\epsilon _{k}(q)(c_{k\uparrow }^{\dagger }c_{k\uparrow
}+c_{k\downarrow }^{\dagger }c_{k\downarrow }).
\end{eqnarray*}%
The $q$-pair expression of $\bar{h}$ is $\bar{h}_{q}=$ $\mathcal{W}_{q}\bar{h%
}\mathcal{W}_{q}^{\dagger }.$

Having common eigenstates $\left\vert \Psi \right\rangle $ of $\tilde{H}_{0}$
and $\bar{h}$, one can obtain the eigenstates $\left\vert \Psi
_{q}\right\rangle $ by the inverse transformation of $\mathcal{W}_{q}$, 
\[
\left\vert \Psi _{q}\right\rangle =\mathcal{W}_{q}\left\vert \Psi
\right\rangle, 
\]%
specifically, replacing all $c_{k\downarrow }$ and $c_{k\downarrow
}^{\dagger }$with $c_{q-k\downarrow }$ $\ $and $c_{q-k\downarrow }^{\dagger
} $ in $\left\vert \Psi \right\rangle $ to obtain $\left\vert \Psi
_{q}\right\rangle$. The corresponding eigenergies $E$ remain unchanged.

The exact solution of $H_{q}$ with $N$ electrons can be written as \cite%
{Richardson}%
\begin{equation}
\left| \Psi _{q}\right\rangle =\prod_{l=1}^{M}S_{l}^{\dagger }\left|
m\right\rangle \text{, }S_{l}^{\dagger }=\sum_{k}\frac{1}{2\epsilon
_{k}(q)-E_{l}}g(k)\eta _{q}^{\dagger }(k),  \label{BCSsolution}
\end{equation}%
where $2M=N-m$ and $m=\sum m_{k}$ is the number of unpaired electrons,
defined by $\eta _{q}(k)\left| m\right\rangle =0$ and $(c_{k\uparrow
}^{\dagger }c_{k\uparrow }+c_{q-k\downarrow }^{\dagger }c_{q-k\downarrow
})\left| m\right\rangle =m_{k}\left| m\right\rangle ,$ in particular $m=0$
for the ground state. It may deserve mentioning that the $m=0$ subspace is
in one to one correspondence with the entire Hilbert space of qubits. It
suggests that a superconductor may act as a natural quantum computer.

\bigskip $E_{l}$ satisfy the Richardson's equation, 
\[
2\sum_{l\neq m}\frac{1}{E_{m}-E_{l}}-\sum_{k}\frac{1-m_{k}}{2\epsilon
_{k}(q)-E_{l}}=\frac{1}{G}, 
\]%
and the eigenenergies of $H_{q}$ are 
\begin{equation}
E=\bar{E}_{q}+\sum_{k}\epsilon _{k}(q)m_{k}+\sum_{l=1}^{M}E_{l},
\label{BCSenergy}
\end{equation}
where $\bar{E}_{q}$ are eigenvalues of $\bar{h}$ and are given by simply
filling the spin-dependent single-particle levels $(\epsilon _{k}-\epsilon
_{k-q})$ for spin-up and $-(\epsilon _{k}-\epsilon _{k-q})$ for spin-down.

The standard BCS treatment may also be interesting in $q$-pairing models
since it is a good approximation for large systems. The BCS approximate
solution is 
\[
E_{BCS}\approx \bar{E}_{q}+2\sum (\epsilon _{k}(q)-\lambda )v_{k}^{2}+\Delta
^{2}/G, 
\]%
where the terms with $v^{4}$ are neglected as usual. The BCS wave function
is 
\[
\left| \Psi _{q}\right\rangle =\prod (u_{k}+v_{k}g(k)\eta _{q}^{\dagger
}(k))\left| 0\right\rangle, 
\]%
which has $d_{x^{2}-y^{2}}$ pairing symmetry when $g(k)$ in eq. (\ref{gk})
is taken \cite{Henley}. The gap parameter $\Delta =\left| G\right| \sum
v_{k}u_{k}$ and 
\[
\left. 
\begin{array}{c}
u_{k}^{2} \\ 
v_{k}^{2}%
\end{array}%
\right\} =\frac{1}{2}\pm \frac{\epsilon _{k}(q)-\lambda }{\sqrt{(\epsilon
_{k}(q)-\lambda )^{2}+\Delta ^{2}}}. 
\]
Again, there are single particle energies $\bar{E}_{q}$, and for the BCS
pairing the energy $\bar{E}_{0}=0$.

The expectation values of an observable are subject to the same dressing
transformations, 
\begin{equation}
\left\langle \Psi _{q}\right| O_{q}\left| \Psi _{q}\right\rangle
=\left\langle \Psi \right| \mathcal{W}_{q}^{\dagger }O_{q}\mathcal{W}%
_{q}\left| \Psi \right\rangle ,  \label{observable}
\end{equation}%
which are well-defined and can be obtained via the known BCS theoretical
methods. There are observables invariant under the dressing transformations,
for instance, the $z$-component of the total spin $\mathcal{W}_{q}^{\dagger
}S_{z}\mathcal{W}_{q}=S_{z}.$ The charge density wave operator $%
Q_{+}=\sum_{k}(c_{q-k\uparrow }^{\dagger }c_{k\uparrow }+c_{q-k\downarrow
}^{\dagger }c_{k\downarrow })$ \cite{SO5} has this property as well, $%
\mathcal{W}_{q}^{\dagger }Q_{+}\mathcal{W}_{q}=Q_{+}$.

\section{Examples and a quantum phase transition}
We would like emphasize that the $\eta $ pairs
or $d$-wave pairs can be treated on the equal footing in this framework, as
long as the different $g(k)$ are used. It is interesting to note for the $%
\eta $ pairs or $d$-wave pairs that $\epsilon _{k}(\pi )=4\epsilon $ if one
uses the single particle levels $\epsilon _{k}=4\epsilon -2\epsilon \cos
k_{x}-2\epsilon \cos k_{y}$ in the two-dimensional attractive Hubbard model.
It will contribute a trivial energy $4\epsilon N$ and will be neglected in
the follow discussions. The solution of $H_{\pi }$ is straightforward since $%
[H_{\pi },G\eta _{\pi }^{\dagger }\eta _{\pi }]=0.$

We consider the one-dimensional case. The ground
state is a result of the competition between the pairing energy 
\[
E_{p}(N)=-UN(2L-N+2)/L, 
\]%
of $G\eta _{\pi }^{\dagger }\eta _{\pi }$ ( $G=-U/L$ is used) and the
single-particle energies $E_{\pi }(N)$ given by $\bar{h}=-2\epsilon
\sum_{k}\cos k(n_{k\uparrow }-n_{k\downarrow })$. The energy before
half-filling is%
\[
\bar{E}_{\pi }(K)=-4\epsilon \frac{\sin \left( K+1\right) a+\sin (Ka)+\sin a%
}{\sin a}, 
\]%
where $a=2\pi /L$ and $N=4K$ since $\bar{h}$ has four-fold degeneracy. The
total energy at half filling is $E_{s}(L)\approx -2\epsilon L/\pi $. After
half-fill it becomes 
\[
\bar{E}_{\pi }(K)=4\epsilon \frac{\sin (\left( K+1\right) a)-\sin (aK)-\sin a%
}{\cos a-1}, 
\]%
where $(N-L)=4K$. Numerical calculation shows that when $U/\epsilon $ is
small, $\bar{h}$ is dominate in the ground state competition, where spins
arranges themselves up and down such that the total magnetization is always
zero for $N=4K$.  We will call this state as MZ independent-particle state or
MZ state. For big values of $U/\epsilon $, the $\pi $ pair (or $d$-wave
pair) correlation $G\eta _{\pi }^{\dagger }\eta _{\pi }$ becomes dominate.
The ground state therefore is in an $\eta _{\pi }^{\dagger }$ condensate, 
\[
\left| \Psi _{\pi }\right\rangle =\frac{1}{\sqrt{N!(L-N)!}}\eta _{\pi
}^{\dagger N}\left| v\right\rangle . 
\]%
When the values of $U/\epsilon $ are inbetween, the ground state is in the
MZ state around half-filling but becomes $\eta _{\pi }^{\dagger }$
condensation as the particle or hole number gets smaller. For instance, when 
$L=10^{4}$, $U/\epsilon <1$ corresponds to the MZ state and $U/\epsilon
>1.31 $ to the $\eta _{\pi }^{\dagger }$ condensation. When $1>U/\epsilon
>1.31,$ the ground states is the MZ states around half-filling and becomes $%
\eta _{\pi }^{\dagger }$ condensation when particle or hole number is small.
Specifically, $U/\epsilon =1.26$, the ground state leaves the MZ state and
becomes $\eta _{\pi }^{\dagger }$ condensation at the doping level $0.2$ and
until full filling.

The single particle energy $\bar{h}$ plays the essential role in this
analysis and the conclusions here are applicable to both the $\eta $ pairs
and $d$-wave pair in two dimensions. (We noticed after finishing the second version
that solutions with $g(k)=1$ are considered in previous works 
\cite{Guan} for the FFLO states).

\section{Extensions.}
It is instructive to rewrite the Hubbard-type
interaction as 
\begin{equation}
V=G\sum_{q=0}\mathcal{W}_{q}\eta _{0}^{\dagger }\eta _{0}\mathcal{W}%
_{q}^{\dagger },  \label{V}
\end{equation}%
where $\mathcal{W}_{0}^{\dagger }=1.$ The expectation values of $V$ under a
wave function $\left| \Phi \right\rangle $ is $\left\langle \Phi \right|
V\left| \Phi \right\rangle =G$Tr$(\eta _{0}^{\dagger }\eta _{0}\rho ),$ where%
\begin{equation}
\rho =\sum_{q=0}\mathcal{W}_{q}^{\dagger }\left| \Phi \right\rangle
\left\langle \Phi \right| \mathcal{W}_{q},  \label{kraus}
\end{equation}%
where $\rho $ is a non-normalized density matrix. This is similar to the
Kraus representation with many \emph{channels}.

The crucial ansatz in the BCS theory is to pick up one pure channel, the BCS
pair \emph{channel} at $q=0$. The ansatz has been verified by numerous
experiments in normal superconductivity. This indicates clearly that the BCS
pairing is dominate in low-lying states in normal
superconductors. Since all pairs in the interaction are on equal footing,
intuitively the single particle energy should be responsible for validity of
the BCS ansatz. In other words, the single particle levels $\epsilon _{k}$
are in favour of the BCS ansatz. Note that the single particle energy $%
h=\sum \epsilon _{k}c_{k\sigma }^{\dagger }c_{k\sigma }$, in the above
discussions, commutes with the total spin $\vec{S}$.

Heretofore, we have not made any physical assumption except generalizing and
studying the BCS ansatz to $q$-pairs. There are evidences that all known
high-temperature superconductors are strongly two-dimensional. Since the $%
q\neq 0$ pairing Hamiltonians keep the conservation of $S_{z}$ and are not
contradict with these evidences, one may extend the forms of the
single-particle levels to spin-dependent ones, for instance considering the
effect from spin-orbital coupling, and writes them as $\sum_{k,\sigma
=\downarrow ,\uparrow }e_{k\sigma }c_{k\sigma }^{\dagger }c_{k\sigma }$. It
and the total $q$-pair Hamiltonian $H_{q}$ commute with $S_{z}$. The
dressing transformation $\mathcal{W}_{q}$ will therefore give 
\begin{eqnarray}
\eta _{q} &\rightarrow &\eta _{0},  \label{correspond} \\
e_{k\uparrow } &\rightarrow &e_{k\uparrow },  \nonumber \\
e_{k\downarrow } &\rightarrow &e_{k-q\downarrow }.  \nonumber
\end{eqnarray}%
In the case that $e_{k\downarrow }=e_{k-q\downarrow }=\epsilon _{k},$ $\eta
_{q}$ will play the exactly same role as $\eta _{0}$. This requires, for the
attractive Hubbard model, $e_{k\uparrow }=4\epsilon -2\epsilon \cos
k_{x}-2\epsilon \cos k_{y}$ and $e_{k\downarrow }=-e_{k\uparrow }$, when $%
q=\pi $. The $\eta $ pairs or $d$-wave pairs are therefore in the same
position as the BCS pairs $\eta _{0}$, while $\eta _{0}$ pairs behave the
same as $\eta _{\pi }$ pairs as discussed in the last section. In this case,
the BCS ansatz implies that the $\eta $ pairs or $d$-wave pairs are dominate
in low-lying states.

Another possible extension is to release the constraints of functions $g(k)$
in the dressing transformations ($\ref{dress1}$). Models generated by any
form of $g(k)$ are still in the exactly solvable family since they can be
unitarily rotated to the BCS model. For instance when $g(k)=2\theta /\pi $
and $q=\pi ,$ a hybrid model 
\begin{eqnarray}
H &=&\cos ^{2}\theta H_{0}+\sin ^{2}\theta H_{\pi }  \label{gtransformation}
\\
&&+\frac{\cos 2\theta }{2}(\sum_{k}\epsilon _{k}c_{q+k\downarrow }^{\dagger
}c_{k\downarrow }+G\eta _{0}^{\dagger }\eta _{\pi }+h.c.),  \nonumber
\end{eqnarray}%
is equivalent to the BCS model and may be used to explain the competition
between the BCS pairs and $\eta $-pairs. Note that now there is a scattering
term between the BCS pairs and $\eta $-pairs. The similar technique is used for quantum state transfer~\cite{WuQST,OH}

Generally, all Hamiltonians $
H=W(H_{0}+h_{B})W^{\dagger }$ with an arbitrary dressing transformation $W$
are in the exactly solvable family, where $h_{B}$ commutes with the BCS
Hamiltonian $H_{0}$ and may even be for another system such as a bath. The
eigenwavefunctions of $H$ are $W\left| \Psi \right\rangle \left|
B\right\rangle $, where $\left| \Psi \right\rangle $ are the BCS
eigenfunctions and if there is a bath, $\left| B\right\rangle $ are eigenfunctions of $h_{B}$.
Eigenvalues are $E+E_{B}$, which are sums of the BCS eigenenergies and
eigenvalues of $h_{B}$. Physically, a BCS pair $\eta _{0}$ now becomes a
dressed BCS pair $W\eta _{0}W^{\dagger }$ and observables have expectation
values with the same form as (\ref{observable}). \emph{All known BCS-type
models can be generated by their own particular dressing transformations }. For
instance, a dressing transformation 
\begin{equation}
W=\exp (i\sum_{k}\phi _{k}c_{-k\downarrow }^{\dagger }c_{-k\downarrow }),
\label{new}
\end{equation}%
can put the BCS pairs into bath, i. e., a phonon bath with $h_{B}=\sum
\omega _{t}b_{t}^{\dagger }b_{t}$, where $\phi _{k}$ are bath operators.
When $\phi _{k}=\sum_{k}\lambda _{k}^{t}b_{t}^{\dagger }b_{t}$, the dressed
Hamiltonian is%
\begin{eqnarray*}
H &=&\sum_{k,\sigma =\downarrow ,\uparrow }\epsilon _{k}n_{k\sigma }+\sum
\omega _{t}b_{t}^{\dagger }b_{t} \\
&&+G\sum_{k,k^{\prime }}e^{i(\phi _{k^{\prime }}-\phi _{k})}\eta
_{0}^{\dagger }(k^{\prime })\eta _{0}(k).
\end{eqnarray*}%
For weaking coupling $\lambda _{k}^{t}\ll 1$, 
\begin{equation}
H\approx H_{0}+h_{B}+iG\sum_{k}\lambda _{k^{\prime }}^{t}b_{t}^{\dagger
}b_{t}\eta _{0}^{\dagger }(k^{\prime })\eta _{0}+h.c,  \label{bath}
\end{equation}
the last term denotes a standard dephasing from the phonon bath. It shows that
the BCS dynamics is naturally fault-tolerant against the dephasing. The method may be 
applicable to energy transport \cite{WD}. 

\section{Conclusion.} 
Using the dressing transformations, we have found an
exactly solvable family of pairing models. The BCS pairs are peculiar in the
family as they live in a three-dimensional spin space, while all other pairs
survives on two dimensional spin space. 
These paring models are on equal footing
but distinguish themselves according to extra single particle energies.
This seems to suggest that a $d$-wave pair is a dressed $s$-wave pair or
vice versa, which might be related to the $d$-wave \cite{Monthoux92} or $s
$-wave \cite{Anderson93} theoretical issue. We look into an example and
notice that the extra energies are responsible for the transitions between
independent-particle states with zero magnetization and collective $\eta $
paring or $d$-wave pairing states. We also secure a condition where another
type of pairs can play the role that the BCS pairs are playing. In addition,
we emphasize that the family of solvable models can be even much bigger.

This work is supported by the Ikerbasque Foundation Start-up,  the Basque Govern- ment (Grant No. IT472-10), the Spanish MINECO/FEDER (No. FIS2012-36673-C03-03), and University of Basque Country UPV/EHU under Program UFI 11/55.

\end{document}